\documentclass[12pt, a4paper] {article}
\usepackage{amsfonts}
\usepackage{verbatim}
\usepackage{latexsym}
\usepackage{amsmath}

\begin{document}

\begin{center}
{\bf \large Canonical brackets from continuous symmetries: Abelian 2-form gauge theory} 

\vskip 1.5cm
{\bf Saurabh Gupta, R. Kumar}\\
{\it Physics Department, Centre of Advanced Studies,}\\
{\it Banaras Hindu University, Varanasi - 221 005, India}\\

\vskip 0.1cm

{\small {e-mails: guptasaurabh4u@gmail.com, raviphynuc@gmail.com}}
\end{center}

\vskip 2 cm

\noindent
{\bf Abstract:} We derive the canonical (anti-)commutation relations amongst the creation 
and annihilation operators of the various basic fields, present in the four $(3 + 1)$ 
- dimensional (4D) {\it free} Abelian 2-from gauge theory, with the help of continuous 
symmetry transformations within the framework of Becchi-Rouet-Stora-Tyutin (BRST) formalism. 
We show that {\it all} the six continuous symmetries of the theory lead to the exactly the same
non-vanishing (anti-)commutator amongst the creation and annihilation operators of the 
normal mode expansion of the basic fields of the theory. \\

\vskip 2.0 cm 



\newpage

\noindent
{\large \bf 1. Introduction}\\

\noindent
Symmetry principles play a crucial role as far as the theoretical description of the 
fundamental interactions of nature is concerned [1]. Three out of four fundamental interactions, 
present in nature (i.e. electromagnetic, weak and strong), are governed by the local 
continuous symmetries known as gauge symmetries. The notion of local gauge invariance 
generates the interaction term for these theories which are endowed with first-class 
constraints in the language of Dirac's prescription for the classification scheme [2]. 

In recent past, the 2-form ($ B^{(2)} = \displaystyle \frac {1} {2!} (dx^\mu 
\wedge dx^\nu) B_{\mu\nu}$) Abelian gauge field $B_{\mu\nu}$ and corresponding gauge 
theories have been studied thoroughly in different contexts [3-5]. Its study is important 
because of its relevance in the context of (super)string 
theories [6,7]. As far as the canonical quantization of such kind of gauge theories are concerned, 
Becchi-Rouet-Stora-Tyutin (BRST) formalism is one of the most intuitive approaches. The BRST 
quantization of such 2-form gauge theories has already been carried out in different contexts [8,9]. 
It has also been shown that the 4D {\it free} Abelian 2-form gauge theory provides a field theoretic 
model for Hodge theory where all the de Rham cohomological operators of the differential geometry 
find their physical realizations in terms of the symmetry transformations (and their 
corresponding generators) of the theory [10]. The quantization of the above Abelian 
2-form gauge theory has been a topic of research interest. The most familiar quantization
scheme is the usual canonical formalism.

In the context of the canonical method of quantization, three steps are followed. 
In the first step, we use the spin-statistics theorem in order to distinguish the nature 
of the fields (i.e. bosonic/fermionic) of a given theory. In the second step, we 
calculate the canonical (graded) Poisson brackets between the field variables and their 
corresponding conjugate momenta and promote them upto the level of (anti-)commutators. In
the third and final step, we express the field variables and corresponding conjugate 
momenta in terms of normal mode expansion of the basic fields, that include the creation 
and annihilation operators. The relevant physical quantities (e.g. conserved charges,
Hamiltonian, etc.) are expressed in terms of creation and annihilation operators where
the concept of normal ordering is required. 

The main motivation of our present investigation is to derive the canonical (anti-)commutation
relations amongst the creation and annihilation operators with the help of continuous 
symmetry transformations (and their corresponding generators) in the context of 4D {\it free}
Abelian 2-form gauge theory. Towards this goal, although, we have taken the help of spin-statistic 
theorem and the concept of normal ordering but we have {\it not} exploited the concept 
of (graded) Poisson brackets. Instead, in the place of latter, we have taken the help of 
continuous symmetry transformations (and their corresponding generators) to obtain the canonical
brackets amongst the creation and annihilation operators. 
It is worthwhile to mention that we have already calculated the canonical brackets 
amongst the creation and annihilation operators with the help of symmetry transformations
in the context of 2D free and 2D interacting Abelian 1-form gauge theory with Dirac fields [11].

Our paper is organized as follows. In the Sec. 2, we discuss the symmetries of the 
4D {\it free} Abelian 2-form gauge theory. Our Sec. 3 contains the conserved charges, that
have been derived with the help of Noether's theorem, and the normal mode expansion of the 
basic fields in terms of the creation and annihilation operators. We have explicitly derived
the (anti-) commutation relations amongst the creation and annihilation operators with the 
help of symmetry principle in Sec. 4 of our manuscript. In Sec. 5, we have calculated, for the 
sake of completeness, the same (anti-)commutation relations with the help of conventional 
Lagrangian formalism. Finally, we have made some concluding remarks in Sec. 6.\\

\noindent
{\large \bf 2. Lagrangian formalism: symmetries}\\

\noindent
We begin with the Becchi-Rouet-Stora-Tyutin (BRST) invariant Lagrangian density 
of a  4D {\it free} Abelian 2-form gauge theory\footnote{We 
choose the flat metric ($\eta_{\mu\nu}$) with the signatures
($ +1, -1, -1, -1$) so that $A \cdot B = \eta_{\mu\nu} A^\mu B^\nu
= A_0 B_0 - A_i B_i $ is the dot product between two non-null vectors
$ A_\mu $ and $B_\mu$ where the Greek indices $\mu, \nu, \eta, . . . . = 0, 1, 2, 3$
and Latin indices $ i, j, k, . . . . = 1, 2, 3$. The  choice of 4D totally antisymmetric Levi-Civita
tensor ($\varepsilon_{\mu\nu\eta\kappa}$) is such that $\varepsilon_{0123}
= +1 = - \varepsilon^{0123},  \varepsilon_{\mu\nu\eta\kappa} \varepsilon^{\mu\nu\eta\kappa} 
= - 4 !$, $\varepsilon_{\mu\nu\eta\kappa}  \varepsilon^{\mu\nu\eta\sigma} = - 3 ! \;
\delta_\kappa^\sigma$, etc. The component $\varepsilon_{0ijk}
= \epsilon_{ijk}$ is the 3D Levi-Civita tensor.}  (see, e.g., [12])

\noindent
\begin{eqnarray}
{\cal L} &=& \frac{1}{2} \; \bigl(\partial^\kappa B_{\kappa \mu} - \partial_\mu \phi_1)
(\partial_\nu B^{\nu \mu} - \partial^\mu \phi_1 \bigr) - \partial_\mu \bar\beta 
\partial^\mu \beta \nonumber\\
&-& \frac{1}{2} \; \bigl(\frac {1}{2}\;  \varepsilon_{\mu\nu\kappa\lambda} \partial^\nu 
B^{\kappa \lambda} - \partial_\mu \phi_2 \bigr) 
\bigl(\frac {1}{2}\;  \varepsilon^{\mu \zeta \sigma\eta} \partial_\zeta  
B_{\sigma \eta } - \partial^\mu \phi_2 \bigr)\nonumber\\
&+& (\partial_\mu \bar C_\nu - \partial_\nu \bar C_\mu) (\partial^\mu C^\nu)
- \frac {1}{2} \; (\partial \cdot \bar C) (\partial \cdot C),
\end{eqnarray}
where $B_{\mu\nu}$ is antisymmetric ($B_{\mu\nu} = - B_{\nu\mu}$) tensor gauge field.  
$\phi_1$ and $\phi_2$ are massless scalar fields, $ (\bar C_\mu) C_\mu$ are the fermionic
(anti-)ghost fields (with $C_\mu^2 = \bar C_\mu^2 = 0, C_\mu \bar C_\nu + \bar C_\nu 
C_\mu = 0$) and $ (\bar \beta) \beta $ are the bosonic (anti-)ghost fields.

The above Lagrangian density remains quasi-invariant under the following on-shell 
nilpotent ($s_{(a)b}^2 = 0$) and absolutely anticommuting $(s_b s_{ab} + s_{ab} s_b = 0)$ 
(anti-)BRST symmetry transformations $s_{(a)b}$ [12]
\begin{eqnarray}
&& s_{ab} B_{\mu\nu} = (\partial_\mu  \bar C_\nu - \partial_\nu \bar C_\mu), 
\quad s_{ab} \bar C_\mu = - \partial_\mu \bar \beta, \quad s_{ab} C_\mu 
= - (\partial^\nu B_{\nu\mu} - \partial_\mu \phi_1), \nonumber\\
&& s_{ab} \phi_1 = \frac{1}{2} \; (\partial \cdot \bar C), \quad s_{ab} \beta 
= - \frac {1}{2} \; (\partial \cdot C), \quad s_{ab} \bar \beta =0, \quad s_{ab} \phi_2 = 0,
\end{eqnarray}
\begin{eqnarray}
&& s_b B_{\mu\nu} = (\partial_\mu  C_\nu - \partial_\nu C_\mu), \quad s_b C_\mu 
= \partial_\mu \beta, \quad s_b \bar C_\mu = (\partial^\nu B_{\nu\mu} 
 - \partial_\mu \phi_1), \nonumber\\
&& s_b \phi_1 = \frac{1}{2} \; (\partial \cdot C), \quad s_b \bar \beta 
= - \frac {1}{2} \; (\partial \cdot \bar C), \quad s_b \beta =0, \quad s_b \phi_2 = 0.
\end{eqnarray}
It can be checked that under the above (anti-)BRST symmetry transformations the kinetic term 
remains invariant. Furthermore, we have following on-shell nilpotent (anti-)co-BRST ($s_{(a)d}^2 = 0$)
symmetry transformations under which the gauge fixing term of the Lagrangian density (1) remains 
invariant. The (anti-)co-BRST symmetry transformations can be given as:
\begin{eqnarray}
&& s_{ad}\; B_{\mu\nu} = \varepsilon_{\mu\nu\kappa\xi} \;\partial^\kappa C^\xi, \qquad s_{ad} 
\;\bar C_\mu = - \Bigl(\frac {1}{2}\; \varepsilon_{\mu\nu\kappa\sigma} \; \partial^\nu 
B^{\kappa\sigma} - \partial_\mu \phi_2 \Bigr),\nonumber\\ 
&& s_{ad}\; {C_\mu} = \partial_\mu \beta, \qquad
s_{ad} \;\phi_2 = \frac {1}{2} \; (\partial \cdot C), \qquad s_{ad} \;\bar \beta 
= - \;\frac {1}{2} \; (\partial \cdot \bar C),\nonumber\\
&& s_{ad} \;(\bar \beta, \; \phi_1, \; \partial^\mu B_{\mu\nu}) = 0,
\end{eqnarray}
\begin{eqnarray}
&& s_d \;B_{\mu\nu} = \varepsilon_{\mu\nu\kappa\xi} \;\partial^\kappa \bar C^\xi, \qquad s_d \;C_\mu 
= \Bigl(\frac {1}{2}\; \varepsilon_{\mu\nu\kappa\sigma} \; \partial^\nu B^{\kappa\sigma} 
- \partial_\mu \phi_2 \Bigr),\nonumber\\ 
&& s_d \;\bar {C_\mu} = - \partial_\mu \bar \beta,\qquad
s_d \;\phi_2 = \frac {1}{2} \; (\partial \cdot \bar C), \qquad s_d \;\beta = - \;\frac {1}{2} 
\; (\partial \cdot C),\nonumber\\
&& s_d \;(\bar \beta, \; \phi_1, \; \partial^\mu B_{\mu\nu}) = 0,
\end{eqnarray} 
these transformations leave the Lagrangian density (1) quasi-invariant. The (anti-)BRST
and (anti-)co-BRST symmetry transformations obey the two key properties
(i) the nilpotency of order two (i.e. $s_{(a)b}^2 = 0, \; s_{(a)d}^2 = 0$), and
(ii) the absolute anticommutativity (i.e. $s_b s_{ab} + s_{ab} s_b = 0,$ \;
$s_d s_{ad} + s_{ad} s_d = 0$). The former property (i.e. nilpotency) shows the 
fermionic nature of (anti-) BRST as well as (anti-)co-BRST symmetries and the latter
property (i.e. absolute anticommutativity) shows that the BRST and anti-BRST 
symmetries are independent of one another (also true in the case of co-BRST
and anti-co-BRST symmetry transformations). 

The anticommutator of the above two symmetries (i.e. $ \{s_b, s_d \} = s_\omega  = - \; 
\{s_{ab}, s_{ad} \} $) lead to a another symmetry, named as bosonic symmetry:
\begin{eqnarray}
s_\omega B_{\mu\nu} = \varepsilon_{\mu\nu\eta\kappa}\; \partial^\eta (\partial_\sigma B^{\sigma\kappa}) 
+ \frac {1}{2}\; \varepsilon_{\nu\xi\sigma\eta}\; \partial_\mu (\partial^\xi B^{\sigma\eta})
- \frac {1}{2}\; \varepsilon_{\mu\xi\sigma\eta}\; \partial_\nu (\partial^\xi B^{\sigma\eta}),\nonumber\\
s_\omega C_\mu = - \frac {1}{2} \partial_\mu (\partial \cdot C), \quad s_\omega \bar C_\mu 
= \frac {1}{2} \partial_\mu (\partial \cdot \bar C), \quad 
s_\omega (\beta, \; \bar \beta, \; \phi_1, \; \phi_2) = 0.
\end{eqnarray}
Under above symmetry transformation the Lagrangian density (1) goes to a total spacetime 
derivative, therefore it is  symmetry of the theory. Moreover, the ghost sector of the 
present theory also possess a continuous symmetry called the ghost symmetry ($s_g$) and 
can be realized as
\begin{eqnarray}
s_g C_\mu = + \Sigma \; C_\mu, \quad s_g \bar C_\mu = - \Sigma \; \bar C_\mu, 
\quad s_g \beta = + 2\Sigma \; \beta, 
\quad s_g \bar \beta = - 2\Sigma \; \bar \beta,
\end{eqnarray}
where $\Sigma$ is a continuous global scale parameter. Thus the 4D {\it free} Abelian 2-form 
gauge theory is endowed with, in totality, six continuous symmetries. \\

\noindent
{\large \bf 3. Conserved charges and normal mode expansions}\\ 

\noindent
We have seen (cf. Sec. 2) the 4D {\it free} Abelian 2-form gauge theory endowed with, in totality, 
six continuous symmetry transformations. According to the Noether's theorem these continuous
symmetries lead to the derivation of six conserved currents ($J^\mu_r, \; r = b, ab, d, ad, g, 
\omega$) corresponding to each symmetry transformations. 
The zeroth component ($J^0_r$) of these conserved currents lead to the following 
conserved charges (i.e. $ Q_r = \int d^3x \; J_r^0 $) 
\begin{eqnarray}
& Q_b = \displaystyle \int d^3x \Bigl [(\partial^0 \bar C^i - \partial^i \bar C^0) (\partial_i \beta)
+ \frac {1}{2} \; (\partial^0 \beta) (\partial \cdot \bar C) - \frac {1}{2}\;
(\partial_i B^{i 0} - \partial^0 \phi_1) (\partial \cdot C)& \nonumber\\
& + H^{0 i j} (\partial_i C_j) - \varepsilon_{ijk}(\partial^i \phi_2)(\partial^j C^k) 
+ (\partial^0 C^i - \partial^i C^0)(\partial^0 B_{0 i} + \partial^j B_{ji} 
- \partial_i \phi_1) \Bigr],&
\end{eqnarray}
\begin{eqnarray}
& Q_{ab} = \displaystyle \int d^3x \Bigl [(\partial^0 C^i - \partial^i C^0) (\partial_i \bar \beta)
+ \frac {1}{2} \; (\partial^0 \bar \beta) (\partial \cdot C) - \frac {1}{2}\;
(\partial_i B^{i 0} - \partial^0 \phi_1) (\partial \cdot \bar C)& \nonumber\\
& + H^{0 i j} (\partial_i \bar C_j) - \varepsilon_{ijk}(\partial^i \phi_2)(\partial^j \bar C^k) 
+ (\partial^0 \bar C^i - \partial^i \bar C^0)(\partial^0 B_{0 i} + \partial^j B_{ji} 
- \partial_i \phi_1) \Bigr],&
\end{eqnarray}
\begin{eqnarray}
Q_d &=& \int d^3x \Bigl[\varepsilon_{ijk} \;(\partial^j \bar C^k)(\partial_0 B^{0i} 
+ \partial_l B^{li} - \partial^i \phi_1 )
+ \frac {1}{2} \; (\partial \cdot C)(\partial_0 \bar \beta )\nonumber\\
&+& \frac {1}{2}\; \Bigl( - \; \varepsilon_{ijk}\; \partial_0 B^{jk} + 2 \;\varepsilon_{ijk}
 \;\partial^j B^{0k}\Bigr) (\partial^i \bar C^0 - \partial^0 \bar C^i)\nonumber\\ 
&+& \frac {1}{4} \; \varepsilon_{ijk}\;(\partial^i B^{jk})(\partial \cdot \bar C)
- \frac {1}{2}\; (\partial \cdot \bar C)(\partial^0 \phi_2 )\nonumber\\
&-& (\partial_i \bar \beta )(\partial^0 C^i - \partial^i C^0) 
+ (\partial_i \phi_2 )(\partial^0 \bar C^i - \partial^i \bar C^0) \Bigr],
\end{eqnarray}
\begin{eqnarray}
Q_{ad} &=& \int d^3x \Bigl[\varepsilon_{ijk} \;(\partial^j C^k)(\partial_0 B^{0i} 
+ \partial_l B^{li} - \partial^i \phi_1 )
+ \frac {1}{2} \; (\partial \cdot \bar C)(\partial_0 \beta )\nonumber\\
&+& \frac {1}{2}\; \Bigl( - \; \varepsilon_{ijk}\; \partial_0 B^{jk} + 2 \;\varepsilon_{ijk}
 \;\partial^j B^{0k}\Bigr) (\partial^i  C^0 - \partial^0 C^i)\nonumber\\ 
&+& \frac {1}{4} \; \varepsilon_{ijk}\;(\partial^i B^{jk})(\partial \cdot C)
- \frac {1}{2}\; (\partial \cdot C)(\partial^0 \phi_2 )\nonumber\\
&-& (\partial_i \beta )(\partial^0 \bar C^i - \partial^i \bar C^0) + (\partial_i \phi_2 )
(\partial^0 C^i - \partial^i C^0) \Bigr],
\end{eqnarray}
\begin{eqnarray}
Q_g &=& \int d^3x \Bigl[(\partial^0 C^i - \partial^i C^0) \bar C_i + (\partial^0 \bar C^i - \partial^i \bar C^0) C_i
- \frac {1}{2} (\partial \cdot C) \bar C_0 \nonumber\\
&-& \frac {1}{2} (\partial \cdot \bar C) C_0 - 2\beta (\partial^0 \bar \beta )
+ 2 \bar \beta (\partial^0 \beta )\Bigr],
\end{eqnarray}
\begin{eqnarray}
Q_\omega &=& \int d^3 x \Bigl [ \frac {1}{2} \; \partial_i (\partial \cdot C) (\partial^0 \bar C^i 
- \partial^i \bar C^0) + \frac {1}{2} \; \partial_i (\partial \cdot \bar C) (\partial^0 C^i 
- \partial^i C^0) \nonumber\\ 
&-& \varepsilon_{ijk}\;  \partial^j (\partial_0 B^{0 k} + \partial_l B^{l k}) (\partial^i \phi_1)
+ \varepsilon_{jlm}\; H^{0ij} \; \partial_i (\partial^l B^{0m})  \nonumber\\
&-& \frac{1}{2} \; \varepsilon_{ijk}\; \partial_0 (\partial^i B^{j k}) (\partial_l B^{l0})
- \frac{1}{2} \; \varepsilon_{ijk}\; \partial_l (\partial^i B^{j k}) (\partial_0 B^{0l} 
+ \partial_m B^{ml}) \nonumber\\
&-& \frac{1}{2} \; \varepsilon_{ijk}\; (\partial_0 B^{jk}) \partial^0 (\partial_0 B^{0i} +
\partial_l B^{li}) + \frac{1}{2} \; \varepsilon_{ijk}\; (\partial^i B^{jk}) \partial_0 (\partial_l B^{l0}) \nonumber\\
&+& \varepsilon_{ijk}\; (\partial^j B^{0k}) \partial^0 (\partial_0 B^{0i} + \partial_l B^{li})
- \frac{1}{2} \; \varepsilon_{jlm}\; H^{0ij} \; \partial_i (\partial^0 B^{lm}) \nonumber\\
&-& \frac{1}{2} \; \varepsilon_{ijk}\; H^{0ij} \; \partial^k (\partial_l B^{l0}) 
+ \frac{1}{2} \; \varepsilon_{ijk}\; H^{0ij} \; \partial_0 (\partial_0 B^{0k} + \partial_l B^{lk}) \nonumber\\
&+& \varepsilon_{ijk}\; \partial^j \big(\partial_0 B^{0 k} + \partial_l B^{l k} \big)
\big(\partial_0 B^{0 i} + \partial_m B^{m i} \big)
+ (\partial_j \phi_2) \; \partial_i H^{0 i j} \Bigr],
\end{eqnarray}
these charges are turn out to be the generators of the corresponding symmetry transformations. 

The Euler-Lagrange equations of motion derived from the Lagrangian density (1) can be given as
\begin{eqnarray}
&& \Box B_{\mu\nu} = 0, \qquad \Box \phi_1 = \Box \phi_2 = 0, 
\qquad \Box \beta = \Box \bar\beta = 0, \nonumber\\
&& \Box C_\mu = \frac {3}{2}\; \partial_\mu (\partial \cdot C), \qquad
\Box \bar C_\mu = \frac {3}{2}\; \partial_\mu (\partial \cdot \bar C),
\end{eqnarray}
we choose the gauge condition such that $(\partial \cdot C) = (\partial \cdot \bar C) = 0$. 
With these gauge conditions the last two equations in (14) reduce to the following form: 
\begin{eqnarray}
\Box C_\mu = 0, \qquad \Box \bar C_\mu = 0. 
\end{eqnarray}

The normal mode expansions of the basic fields, that are present in our theory, in terms 
of the creation and annihilation operators can be given as 
 
\begin{eqnarray}
B_{\mu\nu} (x) &=& \int \frac{d^3x}{\sqrt{(2\pi)^3 \cdot 2 k_0}} \; 
\Big( b_{\mu\nu} (\vec k) \; e^{+ \; i \; k \cdot x} 
+ b^\dagger_{\mu\nu} (\vec k) \; e^{- \; i \; k \cdot x}\Big), \nonumber\\
C_\mu (x) &=& \int \frac{d^3x}{\sqrt{(2\pi)^3 \cdot 2 k_0}} \; 
\Big( c_\mu (\vec k) \; e^{+ \; i \; k \cdot x} 
+ c^\dagger_\mu (\vec k) \; e^{- \; i \; k \cdot x}\Big), \nonumber\\
\bar C_\mu (x) &=& \int \frac{d^3x}{\sqrt{(2\pi)^3 \cdot 2 k_0}} \; 
\Big( \bar c_\mu (\vec k) \; e^{+ \; i \; k \cdot x} 
+ \bar c^\dagger_\mu (\vec k) \; e^{- \; i \; k \cdot x}\Big), \nonumber\\
\beta (x) &=& \int \frac{d^3x}{\sqrt{(2\pi)^3 \cdot 2 k_0}} \; 
\Big(b (\vec k) \; e^{+ \; i \; k \cdot x} 
+ b^\dagger (\vec k) \; e^{- \; i \; k \cdot x}\Big), \nonumber\\
\bar \beta (x) &=& \int \frac{d^3x}{\sqrt{(2\pi)^3 \cdot 2 k_0}} \; 
\Big( \bar b (\vec k) \; e^{+ \; i \; k \cdot x} 
+ \bar b^\dagger (\vec k) \; e^{- \; i \; k \cdot x}\Big), \nonumber\\
\phi_1 (x) &=& \int \frac{d^3x}{\sqrt{(2\pi)^3 \cdot 2 k_0}} \; 
\Big( f_1 (\vec k) \; e^{+ \; i \; k \cdot x} 
+ f^\dagger_1 (\vec k) \; e^{- \; i \; k \cdot x}\Big), \nonumber\\
\phi_2 (x) &=& \int \frac{d^3x}{\sqrt{(2\pi)^3 \cdot 2 k_0}} \; 
\Big( f_2 (\vec k) \; e^{+ \; i \; k \cdot x} 
+ f^\dagger_2 (\vec k) \; e^{- \; i \; k \cdot x}\Big),
\end{eqnarray}
where $k_\mu (= k_0, k_i)$ is the momentum 4-vector and $ b_{\mu\nu}, c_\mu, \bar c_\mu,
b, \bar b, f_1, f_2$ are the annihilation operators and $ b^{\dagger}_{\mu\nu}, c^{\dagger}_\mu,
{\bar c}^{\dagger}_\mu, b^{\dagger}, {\bar b}^{\dagger}, f^{\dagger}_1, f^{\dagger}_2$ are the 
creation operators of the basic fields $ B_{\mu\nu}, C_\mu, \bar C_\mu, \beta, \bar \beta, \phi_1, 
\phi_2$ respectively.

It is interesting to note that the gauge conditions (i.e. $(\partial \cdot C) = (\partial \cdot \bar C) 
= 0 $) imply the following relationships 
\begin{eqnarray}
k^\mu c_\mu (\vec k) \; = \; k^\mu c^\dagger_\mu (\vec k) \; = \; k^\mu \bar c_\mu (\vec k) 
\; = \; k^\mu \bar c^\dagger_\mu (\vec k) \; = \; 0,
\end{eqnarray}
along with  $k^2 = k^\mu k_\mu = 0$. \\

\noindent
{\large \bf 4. Canonical brackets from symmetry principle}\\ 

\noindent
The conserved charges, (cf. (8) - (13)), are turn out to be the generator of the 
continuous symmetry transformations as follows [13,14]
\begin{eqnarray}
s_r \Phi = \pm \;i \; \big [\Phi, \; Q_r \big ]_\pm,\;
\qquad r \; = \; b, \;  ab, \;  d, \; ad, \; \omega, \; g,
\end{eqnarray}
where $\Phi$ is any generic field of the theory and $Q_r$ are the conserved charges.
The $(+)-$ signs, as the subscripts on the square bracket,
correspond to the (anti-)commutators for the generic field $\Phi$ being (fermionic)bosonic 
in nature. The  explanation for the $(+)-$ signs, in front of the square bracket on the r.h.s.
(i.e. $\pm \; i \;\big [\Phi, \; Q_r \big ]_\pm $) is given below:

(i) negative sign is to be taken into account only for the $s_r = s_b, \; s_{ab}, \; s_d, \; s_{ad}$ 
(e.g. $s_b B_{\mu\nu} = - i \; \big [B_{\mu\nu}, \; Q_b \big ],
 \; s_d C_\mu  = - i \; \{ C_\mu, \; Q_d \} $, etc.), and

(ii) for $s_r = s_g, \; s_\omega $ the negative sign is to be taken into account 
only for the bosonic fields and the positive sign is to be chosen for the fermionic 
fields  (e.g. $s_g B_{\mu\nu} = - i\; \big [ B_{\mu\nu}, \; Q_g \big ], 
s_\omega C_\mu = + i \; \big [ C_\mu, \; Q_\omega \big ]$ etc.).

Let us take an example in order to illustrate our formalism;
\begin{eqnarray} 
&&s_d B_{\mu\nu} \; = \; \varepsilon_{\mu\nu\eta\kappa} \; \partial^\eta \bar C^\kappa \; 
= \;  - i \; [ B_{\mu\nu}, Q_d] \Longrightarrow \nonumber\\
&&s_d B_{0i} \; = \; \varepsilon_{ijk} \; \partial^j \bar C^k \; = \;  - i \; [ B_{0i}, Q_d],\nonumber\\
&&s_d B_{ij} \; = \; \varepsilon_{ijk} \; (\partial^0 \bar C^k - \partial^k \bar C^0  )\; 
= \;  - i \; [ B_{ij}, Q_d],
\end{eqnarray}
where $Q_d$ is co-BRST charge and it turns out to be the generator of the 
continuous symmetry transformations $s_d$. Let us choose $ t = 0$ for the sake 
of simplicity in all the computations. Firstly, let us calculate the commutation 
relation corresponding to the $B_{0i}$ component. The l.h.s. of the
second  equation of (19) can be expressed in terms of creation and annihilation 
operators as follows
\begin{eqnarray}
\varepsilon_{ijk}(\partial^j \bar C^k) &=& i \int \frac{d^3 k}{\sqrt {(2\pi)^3\cdot 2k_0}} 
\; \varepsilon_{ijk} \;k^j
\Bigl(\bar c^k (\vec k) \; e^{- i \vec k \cdot \vec x} - (\bar c^k)^\dagger (\vec k) \; 
e^{+ i \vec k \cdot \vec x}\Bigr)\nonumber\\
&\equiv &  \frac{i}{2}\; \int \frac{d^3 k}{\sqrt {(2\pi)^3\cdot 2k_0}} \; \varepsilon_{ijk} \;k^j
\Bigl[ \Bigl(\bar c^k (\vec k) \; e^{- i \vec k \cdot \vec x} - (\bar c^k)^\dagger (\vec k) 
\; e^{+ i \vec k \cdot \vec x}\Bigr) \nonumber\\
&+& \Bigl(\bar c^k (\vec k) \; e^{- i \vec k \cdot \vec x} - (\bar c^k)^\dagger (\vec k) 
\; e^{+ i \vec k \cdot \vec x}\Bigr)\Bigr].
\end{eqnarray}
The reason for breaking it into two similar terms will be clear later when we compare the exponentials.
In the second term of the above equation, changing $ \vec k \to -\vec k$ and rearranging the terms, we obtain 
\begin{eqnarray} 
\varepsilon_{ijk}(\partial^j \bar C^k) &=& \frac{i}{2}\; \int \frac{d^3 k}{\sqrt {(2\pi)^3\cdot 2k_0}} 
\; \varepsilon_{ijk} \;k^j 
\Big[ \Big(\bar c^k (\vec k) + (\bar c^k)^\dagger (-\vec k)\Big)\; e^{-i\vec k \cdot \vec x}\nonumber\\
&-& \Big(\bar c^k (-\vec k) + (\bar c^k)^\dagger (\vec k)\Big)\; e^{+i\vec k \cdot \vec x} \Big].
\end{eqnarray}
Now the r.h.s. of equation (19), we re-express $B_{0i}(\vec x)$ in terms of creation and annihilation operators
as follows (at $t=0$):
\begin{eqnarray}
&& - i [B_{0i} (\vec x), \; Q_d] = -i \Big[\int \frac{d^3k}{\sqrt{(2\pi)^3 \cdot 2k_0}} \; \Big( b_{0i} 
(\vec k)\; e^{-i\vec k \cdot \vec x}  + b^\dagger_{0i} (\vec k) \;e^{+i\vec k \cdot \vec x}\Big), \; Q_d \Big]\nonumber\\
&& = -i \int \frac{d^3k}{\sqrt{(2\pi)^3 \cdot 2k_0}} \Big( [b_{0i} (\vec k),\; Q_d]\; e^{-i \vec k \cdot \vec x} 
\; + \; [b^\dagger_{0i} (\vec k),\; Q_d]\; e^{+i \vec k \cdot \vec x} \Big).
\end{eqnarray}
Comparing the exponentials from the r.h.s. of the equations (21) and (22), we obtain following 
relationship 
\begin{eqnarray}
 \; [b_{0i} (\vec k), \; Q_d] &=& - \frac{1}{2} \;\varepsilon_{ijk} \; k^j \Big( \bar c^k (\vec k) 
+ (\bar c^k)^\dagger (-\vec k)  \Big) \nonumber\\
&\equiv& - \int \frac {d^3p}{2} \;\varepsilon_{ijk} \; p^j \Big( \bar c^k (\vec p) + (\bar c^k)^\dagger (-\vec p)  
\Big) \;\delta^{(3)}(\vec k - \vec p),
\end{eqnarray}
\begin{eqnarray}
 \; [b^\dagger_{0i} (\vec k), \; Q_d] &=& \frac{1}{2} \;\varepsilon_{ijk} \; k^j \Big( \bar c^k (-\vec k) 
+ (\bar c^k)^\dagger (\vec k)  \Big) \nonumber\\
&\equiv& \int \frac {d^3p}{2} \; \varepsilon_{ijk} \; p^j \Big( \bar c^k (-\vec p) + (\bar c^k)^\dagger (\vec p)  
\Big) \;\delta^{(3)}(\vec k - \vec p). 
\end{eqnarray}
The relevant part of $Q_d$ (that have non-vanishing (anti-)commutation relations with $ B_{0 i} (\vec x)$) 
can be given as (cf. (10))
\begin{eqnarray}
Q_d \approx  \int d^3 y\; \varepsilon_{ljk}\; (\partial^j \bar C^k)(\partial_0 B^{0l}).
\end{eqnarray}
The above equation for $Q_d$ can be re-expressed in terms of the normal mode expansions of the basic 
fields (cf. (16))
\begin{eqnarray}
Q_d &\approx &  \int \frac {d^3 y\;d^3 p\;d^3 q}{(2\pi )^3 \cdot \sqrt {2p_0 \cdot 2q_0 }}
\;\varepsilon_{ljk} (- p^j q_0)
\Big( \bar c^k (\vec p) \; b^{0l} (\vec q) \;e^{-\; i\;(\vec p + \vec q)\cdot \vec y} \nonumber\\
&-& \bar c^k (\vec p) \; (b^{0l})^\dagger (\vec q) \;e^{-\; i\;(\vec p - \vec q)\cdot \vec y}
- (\bar c^k)^\dagger (\vec p) \; b^{0l} (\vec q) \;e^{+\; i\;(\vec p - \vec q)\cdot \vec y}\nonumber\\
&+& (\bar c^k)^\dagger (\vec p)\; (b^{0l})^\dagger (\vec q) \;e^{+\; i\;(\vec p + \vec q)\cdot \vec y}\Big).
\end{eqnarray}
Integrating\footnote{We have used the following definition of Dirac $\delta$-function\\
$\displaystyle\int d^3x \; e^{\pm i (\vec p - \vec q) \cdot \vec x} 
= (2\pi)^3 \;\delta^{(3)} (\vec p - \vec q), \quad \displaystyle\int d^3q \; f(\vec q) 
\;\delta^{(3)} (\vec p - \vec q) = f(\vec p).$}
the above equation with respect to $d^3 y$ and $d^3 q$
\begin{eqnarray}
Q_d &\approx & - \frac {1}{2} \int d^3p \; (\varepsilon_{ljk} \;p^j ) \Big( \bar c^k (\vec p) \;b^{0l} (-\vec p) 
+ (\bar c^k)^\dagger (\vec p)\;( b^{0l})^\dagger (-\vec p) \nonumber\\
&-& (\bar c^k)^\dagger (\vec p)\; b^{0l} (\vec p) - \bar c^k (\vec p) \;(b^{0l})^\dagger (\vec p) \Big).
\end{eqnarray}
In the first two terms of above equation changing $\vec p \to - \vec p$ \;and collecting the 
coefficient of $b^{0l}(\vec p)$ and $(b^{0l})^\dagger (\vec p)$, we obtain
\begin{eqnarray}
Q_d &\approx &  \frac {1}{2} \int d^3p \; (\varepsilon_{ljk} \;p^j ) \Big[ \Big( (\bar c^k)^\dagger (\vec p) 
+ \bar c^k (-\vec p)\Big)\; b^{0l}(\vec p) \nonumber\\
&+& \Big((\bar c^k)^\dagger (- \vec p) + \bar c^k (\vec p)\Big) \;(b^{0l})^\dagger (\vec p) \Big].
\end{eqnarray}
Now substituting this value of $Q_d$ in expression (23), we have following 
expression that contain the two existing commutators
\begin{eqnarray}
[b_{0i} (\vec k), \; Q_d] &=&  \frac {1}{2} \int d^3p \; (\varepsilon_{ljk} \;p^j ) 
\Big( \big( (\bar c^k)^\dagger (\vec p) 
+ \bar c^k (-\vec p)\big)\; [b_{0i}(\vec k),\; b^{0l}(\vec p)] \nonumber\\
&+& \big( \bar c^k ( \vec p) + (\bar c^k)^\dagger (- \vec p)\big) \;[b_{0i} (\vec k),\;
(b^{0l})^\dagger (\vec p)] \Big).
\end{eqnarray}
In fact, we have eight commutators out of which four commutators, amongst the (anti-)ghost fields with 
the rest of the bosonic fields, are zero because the (anti-)ghost fields are decoupled from 
the rest of the bosonic fields present in the theory. Now
comparing the r.h.s. of (23) and (29), we obtain following commutators
\begin{eqnarray}
\big[ b_{0i} (\vec k), \; b^{0l} (\vec p)\big] = 0, \qquad \big[ b_{0i} (\vec k), 
\; (b^{0l})^\dagger (\vec p)\big] = - \; \delta_i^l \; \delta^{(3)}(\vec k - \vec p).
\end{eqnarray}
It is straightforward to check that, if we substitute the value of $Q_d$ (cf. (28)) in the 
equation (24), in stead of equation (23), we get the exactly same set of 
canonical commutation relations (cf. (30)).

Secondly, let us calculate the commutation relations corresponding to the $B_{ij}$ component (cf. (19))
\begin{eqnarray}
s_d B_{ij}  = \; \varepsilon_{ijl} \; (\partial^0 \bar C^l - \partial^l \bar C^0)
= \; - i \; [B_{ij}, \; Q_d] \;.
\end{eqnarray}
The l.h.s. of above equation can be expanded in the terms of creation and annihilation operators in the 
following manner 
\begin{eqnarray}
\varepsilon_{ijl} (\partial^0 \bar C^l - \partial^l \bar C^0) &=& i \int \frac {d^3k}{\sqrt{(2\pi)^3 \cdot 2k_0}} \; 
\varepsilon_{ijl} \Big[ \Big(k^0 \bar c^l (\vec k) - k^l \bar c^0 (\vec k) \Big) e^{- i \vec k \cdot \vec x} \nonumber\\
&-& \Big(k^0 (\bar c^l)^\dagger (\vec k) - k^l (\bar c^0)^\dagger (\vec k) \Big) e^{+ i \vec k \cdot \vec x}
\Big]\nonumber\\
&\equiv & \frac{i}{2} \int \frac {d^3k}{\sqrt{(2\pi)^3 \cdot 2k_0}} \; 
\varepsilon_{ijl} \Big[ \Big(k^0 \bar c^l (\vec k) - k^0 (\bar c^l)^\dagger (-\vec k) \nonumber\\
&-& k^l \bar c^0 (\vec k) - k^l (\bar c^0)^\dagger (-\vec k)  \Big) e^{- i \vec k \cdot \vec x} 
- \Big(k^0 (\bar c^l)^\dagger (\vec k) - k^0 \bar c^l (- \vec k)\nonumber\\
&-& k^l (\bar c^0)^\dagger (\vec k) - k^l \bar c^0 (-\vec k)  \Big) e^{+ i \vec k \cdot \vec x} \Big].
\end{eqnarray}
In the r.h.s. of equation (31), we re-express $B_{ij} (\vec x)$ in terms of creation and 
annihilation operators as follows
\begin{eqnarray}
&&- i [B_{ij} (\vec x), \; Q_d] = -i \Big[\int \frac{d^3k}{\sqrt{(2\pi)^3 \cdot 2k_0}} \; \Big( b_{ij} (\vec k)
\; e^{-i\vec k \cdot \vec x}
+ b^\dagger_{ij} (\vec k) \;e^{+i\vec k \cdot \vec x}\Big), \; Q_d \Big]\nonumber\\
&&= -i \int \frac{d^3k}{\sqrt{(2\pi)^3 \cdot 2k_0}} \Big( [b_{ij} (\vec k),\; Q_d]\; e^{-i \vec k \cdot \vec x} 
+ [b^\dagger_{ij} (\vec k),\; Q_d]\; e^{+i \vec k \cdot \vec x} \Big). 
\end{eqnarray}
Now comparing the exponentials from the equations (32) and (33), we get
\begin{eqnarray}
\big[b_{ij} (\vec k),\; Q_d\big] &=& - \frac {1}{2}\;\varepsilon_{ijl} \;\Big(k^0 \bar c^l (\vec k) 
- k^0 (\bar c^l)^\dagger (-\vec k) 
- k^l \bar c^0 (\vec k) - k^l (\bar c^0)^\dagger (-\vec k)  \Big) \nonumber\\
&\equiv& - \frac{1}{2} \; \int d^3p \; \varepsilon_{ijl} \; \delta^{(3)}(\vec k 
- \vec p) \; \Big( p^0 \; \bar c^l (\vec p) 
- p^0 \; (\bar c^l)^\dagger (- \vec p) \nonumber\\
&-& p^l \; \bar c^0 (\vec p) - p^l (\bar c^0)^\dagger (-\vec p)\Big),
\end{eqnarray}
\begin{eqnarray}
\big[ b^\dagger_{ij} (\vec k),\; Q_d\big] &=& \frac {1}{2}\;\varepsilon_{ijl}\Big(k^0 
(\bar c^l)^\dagger (\vec k) - k^0 \bar c^l (- \vec k)
- k^l (\bar c^0)^\dagger (\vec k) - k^l \bar c^0 (-\vec k)  \Big) \nonumber\\
&\equiv& \frac{1}{2} \; \int d^3p \; \varepsilon_{ijl} \; \delta^{(3)}(\vec k - \vec p) \; \Big( p^0 \;
(\bar c^l)^\dagger (\vec p) - p^0 \; \bar c^l (- \vec p) \nonumber\\
&-& p^l \; (\bar c^0)^\dagger (\vec p) - p^l \; \bar c^0 (-\vec p) \Big).
\end{eqnarray}
The relevant part of $Q_d$ (that have non-vanishing commutation relations with $B_{ij} (\vec x)$)
can be given as (cf. (10))
\begin{eqnarray}
Q_d &\approx & - \frac {1}{2}\int d^3y \; \varepsilon_{lmn} (\partial^0 B^{mn}) (\partial^l \bar C 
- \partial^0 \bar C^l),
\end{eqnarray}
re-expressing $Q_d$ in terms of normal mode expansions of basic fields (cf. (16)): 
\begin{eqnarray}
Q_d &\approx & -\;\frac {1}{2} \int \frac {d^3 y \;d^3 p\;d^3 q}{(2\pi )^3\cdot \sqrt{2p_0 \cdot 2q_0}} 
\; \varepsilon_{lmn}\; p^0
\Big[\Bigl(q^0 \;b^{mn} (\vec p) \; \bar c^l (\vec q)\nonumber\\
&-& q^l \;b^{mn} (\vec p) \; \bar c^0 (\vec q)\Big)
\;e^{- i (\vec p + \vec q)\cdot \vec y} \nonumber\\
&-& \Big(q^0 \;b^{mn} (\vec p) \; (\bar c^l)^\dagger (\vec q) - q^l \;b^{mn} (\vec p) \; 
(\bar c^0)^\dagger (\vec q)\Big)\;e^{- i (\vec p - \vec q)\cdot \vec y}\nonumber\\
&-& \Big( q^0 \;(b^{mn})^\dagger (\vec p) \; \bar c^l (\vec q) - q^l \;(b^{mn})^\dagger (\vec p) \;
\bar c^0 (\vec q)\Big) \; e^{+ i(\vec p - \vec q)\cdot \vec y}\nonumber\\
&+& \Big( q^0 \;(b^{mn})^\dagger (\vec p) \; (\bar c^l)^\dagger (\vec q) - q^l \;(b^{mn})^\dagger 
(\vec p) \; (\bar c^0)^\dagger (\vec q)\Big) \; e^{+ i (\vec p + \vec q)\cdot \vec y} \Big].
\end{eqnarray}
Integrating out $d^3y$ and $d^3q$ from the above equation, we obtain 
\begin{eqnarray}
Q_d &\approx & -\;\frac {1}{2} \int \frac {d^3 p}{2} \; \varepsilon_{lmn}
\Big[p^0 \;b^{mn} (\vec p) \Big(\bar c^l (-\vec p) - (\bar c^l)^\dagger (\vec p) \Big) \nonumber\\
&-& p^0 \;(b^{mn})^\dagger (\vec p) \Big( \bar c^l (\vec p) - (\bar c^l)^\dagger (-\vec p)\Big)
+ p^l \;b^{mn}(\vec p) \Big(\bar c^0 (-\vec p) + (\bar c^0)^\dagger (\vec p)\Big) \nonumber\\
&+& p^l \;(b^{mn})^\dagger (\vec p) \Big( \bar c^0 (\vec p) + (\bar c^0)^\dagger (-\vec p)\Big) \Big].
\end{eqnarray}
Now substituting this value of $Q_d$ in the l.h.s. of expression (34), we have following expression
having four commutators 
\begin{eqnarray}
\big[b_{ij} (\vec k),\; Q_d\big] &=& - \frac {1}{2} \int \frac {d^3p}{2} \;\varepsilon_{lmn} 
\Big(p^0 [b_{ij}(\vec k), \;b^{mn} (\vec p)] \big(\bar c^l (-\vec p) - (\bar c^l)^\dagger 
(\vec p) \big) \nonumber\\
&-& p^0 [b_{ij} (\vec k), \;(b^{mn})^\dagger (\vec p)] \big( \bar c^l (\vec p) - (\bar c^l)^\dagger 
(-\vec p)\big)\nonumber\\
&+& p^l [b_{ij} (\vec k), \;b^{mn}(\vec p)] \big(\bar c^0 (-\vec p) + (\bar c^0)^\dagger 
(\vec p)\big)\nonumber\\ 
&+&  p^l [b_{ij} (\vec k), \; (b^{mn})^\dagger (\vec p)] \big( \bar c^0 (\vec p) + (\bar c^0)^\dagger 
(-\vec p)\big) \Big),
\end{eqnarray}
In fact there exist, in totality,  sixteen commutators out of which  eight commutators (that involve the 
(anti-)ghost fields) are zero because of the fact that
the (anti-)ghost fields are decoupled from the rest of the bosonic fields present in the
theory. Hence, the (anti-)commutators of the (anti-)ghosts fields with the rest of all the 
bosonic fields are zero. Now rearranging the various terms, we have following expression 
\begin{eqnarray}
\big[b_{ij} (\vec k),\; Q_d\big] &=& - \frac {1}{2} \int \frac {d^3p}{2} \;\varepsilon_{lmn} 
\Big([b_{ij}(\vec k), \;b^{mn} (\vec p)] \big(p^0\;\bar c^l (-\vec p) - p^0\;(\bar c^l)^\dagger 
(\vec p) \nonumber\\
&+& p^l \; \bar c^0 (-\vec p) + p^l \;(\bar c^0)^\dagger (\vec p)\big) \nonumber\\
&-& [b_{ij} (\vec k), \;(b^{mn})^\dagger (\vec p)] \big( p^0 \;\bar c^l (\vec p) - p^0 
\;(\bar c^l)^\dagger (-\vec p) -  p^l\;\bar c^0 (\vec p)\nonumber\\
&-& p^l\; (\bar c^0)^\dagger (-\vec p)\big) \Big).
\end{eqnarray}
Comparing the r.h.s. of (34) and (40), we get the following commutation relations amongst the
creation and annihilation operators 
\begin{eqnarray}
[b_{ij}(\vec k), \;b^{mn} (\vec p)] = 0, \quad [b_{ij} (\vec k), \;(b^{mn})^\dagger (\vec p)] = 
- (\delta_i^m \delta_j^n- \delta_i^n \delta_j^m) \delta^{(3)}(\vec k - \vec p).
\end{eqnarray}
It is worthwhile to mention that, instead of substituting the value of $Q_d$ (cf. (38)) into 
equation (34), if we substitute its value into equation (35), we get exactly same set of 
canonical commutation relations as in (41).

Similar exercise can also be done with the other basic fields of the present theory. It is 
straightforward to check that, with the help of symmetry principle, we get the same set of 
(anti-)commutation relations amongst the creation and annihilation operators as obtained 
from the conventional Lagrangian formalism (cf. (49) below). 
It is worthwhile to mention that {\it all} the six continuous symmetries of the present 4D {\it free}
Abelian 2-form gauge theory lead to the exactly same set of (anti-)commutation relations amongst the 
creation and annihilation operators (cf. (49) below).\\

\noindent
{\large \bf 5. Canonical brackets from Lagrangian formalism }\\

\noindent
It is evident to calculate the canonical conjugate momenta form the Lagrangian (1) and 
are listed below;
\begin{eqnarray}
&& \Pi_{(\phi_1)} = \dot \phi_1 + \partial_i B^{0i}, \quad \Pi_{(\phi_2)} = - \dot \phi_2
+ \frac {1}{2} \; \varepsilon^{ijk} \; \partial_i B_{jk}, 
\quad \Pi_{(\beta)} = - \dot {\bar\beta}, \nonumber\\
&& \Pi_{(\bar \beta)} = - \dot \beta, \qquad \Pi^0_{(C)} = \frac {1}{2} \; (\partial \cdot \bar C),
\qquad \Pi^i_{(C)} = - (\partial^0 \bar C^i - \partial^i \bar C^0), \nonumber\\
&& \Pi^i_{(\bar C)} = (\partial^0  C^i - \partial^i C^0),
\quad \Pi^{i j}_{(B)} = \frac {1}{2}\;  H^{0 i j} 
- \frac {1}{2} \; \varepsilon^{i j k} (\partial_k \phi_2),\nonumber\\
&& \Pi^0_{(\bar C)} = - \frac {1}{2} \; (\partial \cdot C), \quad 
\Pi^{0 i}_{(B)} = \frac {1}{2} \; (\partial_0 B^{0 i} + \partial_j B^{j i} 
- \partial^i \phi_1),
\end{eqnarray}
here we have adopted the left derivative prescription for the fermionic fields.
Therefore, the canonical (anti-)commutation relations amongst the basic fields of the 
theory and corresponding canonically conjugate momenta can be given as (at equal time $t$)
\begin{eqnarray}
&& [ \phi_1 (\vec x,t), \;\dot \phi_1 (\vec y, t) ] \; 
= \; i \; \delta^{(3)} (\vec x - \vec y), \quad
[ \phi_2 (\vec x,t),\; \dot \phi_2 (\vec y,t) ] \; 
=  - i \; \delta^{(3)} (\vec x - \vec y), \nonumber\\
&& [ \beta (\vec x,t), \;\dot {\bar \beta} (\vec y,t) ] \; 
= \; - i \; \delta^{(3)} (\vec x - \vec y), \quad
 [ \bar \beta (\vec x,t),\; \dot \beta (\vec y,t) ] \; 
= \; - i \; \delta^{(3)} (\vec x - \vec y), \nonumber\\
&& \{ C_0 (\vec x,t), \; \dot {\bar C}^0 (\vec y,t) \} \;
= \; 2 i \; \delta^{(3)} (\vec x - \vec y), \nonumber\\
&& \{ C_i (\vec x,t),\; \dot {\bar C}^j (\vec y,t) \} \;
= - i \; \delta_i^j \; \delta^{(3)} (\vec x - \vec y), \nonumber\\
&& \{ \bar C_0 (\vec x,t),\; \dot C^0 (\vec y,t) \} \;
= - 2 i \; \delta^{(3)} (\vec x - \vec y), \nonumber\\
&&\{ \bar C_i (\vec x,t),\; \dot C^j (\vec y,t) \} \;
= \; i \; \delta_i^ j \; \delta^{(3)} (\vec x - \vec y), \nonumber\\
&& [ B_{0 i} (\vec x,t), \dot B^{0 j} (\vec y,t) ] \;
=  i \; \delta^j_i \; \delta^{(3)} (\vec x - \vec y), \nonumber\\
&&[B_{i j} (\vec x,t), \dot B^{k l} (\vec y,t)] \; 
=  i \; (\delta^k_i \delta^l_j - \delta^k_j \delta^l_i) \;\delta^{(3)} (\vec x - \vec y).
\end{eqnarray}
All the rest of (anti-)commutators are zero. Now our aim is to calculate the (anti-)commutation relations
amongst the creation and annihilation operators of the theory. In order to simplify our computations, we re-express
the normal mode expansions of the basic fields (cf. (16)) according to [13]

\begin{eqnarray*}
B_{\mu\nu}(\vec x,t) &=& \int d^3x \Big( f_k^* (x) \; b_{\mu\nu} (k) + f_k (x) 
\; b_{\mu\nu}^\dagger (k) \Big),\nonumber\\
C_\mu(\vec x,t) &=& \int d^3x \Big( f_k^* (x) \; c_\mu (k) + f_k (x) \; c_\mu^\dagger (k) \Big),\nonumber\\
\bar C_\mu(\vec x,t) &=& \int d^3x \Big( f_k^* (x) \; \bar c_\mu (k) + f_k (x) \; 
\bar c_\mu^\dagger (k) \Big),\nonumber\\
\beta (\vec x,t) &=& \int d^3x \Big( f_k^* (x) \; b(k) + f_k (x) \; b^\dagger (k) \Big),
\end{eqnarray*}
\begin{eqnarray}
\bar \beta (\vec x,t) &=& \int d^3x \Big( f_k^* (x) \; \bar b(k) + f_k (x) \; \bar b^\dagger (k) \Big),\nonumber\\
\phi_1(\vec x,t) &=& \int d^3x \Big( f_k^* (x) \; f_1(k) + f_k (x) \; f_1^\dagger (k) \Big),\nonumber\\
\phi_2(\vec x,t) &=& \int d^3x \Big( f_k^* (x) \; f_2(k) + f_k (x) \; f_2^\dagger (k) \Big),
\end{eqnarray}
where $f_k (x)$ and $f_k^* (x)$ are defined as 
\begin{eqnarray}
f_k (x) = \frac {e^ {-i k \cdot x}} {\sqrt {(2 \pi)^3 \; 2 k_0}}, \qquad
f^*_k (x) = \frac {e^ {i k \cdot x}} {\sqrt {(2 \pi)^3 \; 2 k_0}}.
\end{eqnarray}
These functions (i.e. $f_k (x)$ and $f_k^* (x)$) form an orthonormal set. It is self-evident 
from the following conditions
\begin{eqnarray}
&& \int d^3x \;  f^*_k (x) \; i \overleftrightarrow{\partial_0} \; f_{k'} (x) 
= \delta^{(3)} (\vec k - \vec k'), \nonumber\\
&& \int d^3x \; f^*_k (x) \; i \overleftrightarrow{\partial_0} \; f^*_{k'} (x) = 0, \quad
\int dx \; f_k (x) \; i \overleftrightarrow{\partial_0} \; f_{k'} (x) = 0.
\end{eqnarray} 
In the above equation, the following standard definition of operator $\overleftrightarrow{\partial_0}$has been taken into account
\begin{eqnarray}
A \;  \overleftrightarrow{\partial_0} \; B = A (\partial_0 B) - (\partial_0 A) B. 
\end{eqnarray}
Using the above relationships, we express the creation and annihilation operators 
in terms of the basic fields and the orthonormal functions (i.e. $f_k (x)$ and $f_k^* (x)$)
as follows:

\begin{eqnarray}
b_{\mu\nu} (k) &=& \int d^3x \; B_{\mu\nu} (x) \;i \overleftrightarrow {\partial_0} \; f_k (x), \quad
b^\dagger_{\mu\nu} (k) = \int d^3x \; f^*_k (x) \; i \overleftrightarrow {\partial_0} \; B_{\mu\nu} (x), \nonumber\\
c_\mu (k) &=& \int d^3x \; C_\mu (x) \;i \overleftrightarrow {\partial_0} \; f_k (x), \quad
c^\dagger_\mu (k) = \int d^3x \; f^*_k (x) \; i \overleftrightarrow {\partial_0} \; C_\mu (x), \nonumber\\
\bar c_\mu (k) &=& \int d^3x \; \bar C_\mu (x) \;i \overleftrightarrow {\partial_0} \; f_k (x), \quad
\bar c^\dagger_\mu (k) = \int d^3x \; f^*_k (x) \; i \overleftrightarrow {\partial_0} \; \bar C_\mu (x), \nonumber\\
b(k) &=& \int d^3x \; \beta (x) \;i \overleftrightarrow {\partial_0} \; f_k (x), \quad
b^\dagger (k) = \int d^3x \; f^*_k (x) \; i \overleftrightarrow {\partial_0} \; \beta (x), \nonumber\\
\bar b(k) &=& \int d^3x \; \bar \beta (x) \;i \overleftrightarrow {\partial_0} \; f_k (x),\quad
\bar b^\dagger (k) = \int d^3x \; f^*_k (x) \; i \overleftrightarrow {\partial_0} \; \bar \beta (x), \nonumber\\
f_1 (k) &=& \int d^3x \; \phi_1 (x) \;i \overleftrightarrow {\partial_0} \; f_k (x), \quad
f^\dagger_1 (k) = \int d^3x \; f^*_k (x) \; i \overleftrightarrow {\partial_0} \; \phi_1 (x), \nonumber\\
f_2(k)& =& \int d^3x \; \phi_2(x) \;i \overleftrightarrow {\partial_0} \; f_k (x), \quad
f^\dagger_2 (k) = \int d^3x \; f^*_k (x) \; i \overleftrightarrow {\partial_0} \; \phi_2 (x).
\end{eqnarray}
Now it is straight forward to check, using above relationships (48),  that we have following 
non-vanishing (anti-)commutation relations amongst the creation and annihilation operators of 
the 4D {\it free} Abelian 2-form gauge theory  
\begin{eqnarray}
&& [ b(\vec k), \bar b^\dagger (\vec k^\prime) ] \; 
= \; \delta^{(3)} (\vec k - \vec k^\prime), \quad
[ \bar b(\vec k), b^\dagger (\vec k^\prime) ] \; 
= \; \delta^{(3)} (\vec k - \vec k^\prime), \nonumber\\
&& [ f_1(\vec k), f_1^\dagger (\vec k^\prime) ] \; 
= - \delta^{(3)} (\vec k - \vec k^\prime), \quad
\{ c_0(\vec k), (\bar c^0)^\dagger (\vec k^\prime) \} \; 
= - 2 \; \delta^{(3)} (\vec k - \vec k^\prime), \nonumber\\
&& \{ \bar c_0(\vec k),  (c^0)^\dagger (\vec k^\prime) \} \; 
= 2 \; \delta^{(3)} (\vec k - \vec k^\prime),\quad
\{ c_i^\dagger (\vec k), \bar c^j(\vec k^\prime) \} 
= - \; \delta_i^j \; \delta^{(3)} (\vec k - \vec k^\prime), \nonumber\\
&& \{ c_i(\vec k),  (\bar c^j)^\dagger (\vec k^\prime) \} \; 
= \; \delta_i^j \; \delta^{(3)} (\vec k - \vec k^\prime), \quad [ f_2(\vec k), f_2^\dagger (\vec k^\prime) ] \; 
= \delta^{(3)} (\vec k - \vec k^\prime),\nonumber\\
&& [ b_{0 i} (\vec k),( b^{0 j})^\dagger (\vec k^\prime) ] \;
=  - \; \delta_i^j \; \delta^{(3)} (\vec k - \vec k^\prime), \nonumber\\
&&[ b_{i j } (\vec k), (b^{m n})^\dagger (\vec k^\prime) ] \; 
= - \; (\delta_i^m \delta_j^n - \delta_i^n \delta_j ^m)  
\; \delta^{(3)} (\vec k - \vec k^\prime).
\end{eqnarray}
All the rest of the (anti-)commutators amongst the creation and annihilation operators are zero. \\

\noindent
{\large \bf 6. Conclusion}\\

\noindent
We have derived all the (anti-)commutation relations amongst the creation and annihilation
operators of 4D {\it free} Abelian 2-form gauge theory with the help of symmetry 
principle. The key feature of our present investigation is that while deriving the 
(anti-)commutation relations amongst the creation and annihilation operators, although 
we have taken the help of spin-statistics theorem and normal ordering but we have {\it not}
used the concept of (graded) Poisson brackets. Instead of latter, we have taken the 
help of continuous symmetry transformations present in the theory. All the six continuous 
symmetries (and their corresponding generators) present in the theory lead to the 
same set of (anti-)commutation relations amongst the creation and annihilation operators. 
This is a unique feature of our present investigation.\\

\noindent
{\large \bf Acknowledgments}\\ 

\noindent
We would like to thank R. P. Malik for introducing us to this problem. We both (i.e. SG and RK)
would like to thank CSIR and UGC, New Delhi, respectively, for financial support.\\

\end{document}